\documentclass{iau_FM}
\usepackage{graphicx}

\title[Multiple Solar Jets from NOAA AR 12644] 
{Multiple Solar Jets from NOAA AR 12644}
\author[Reetika Joshi \& Ramesh Chandra]   
{Reetika Joshi
\and Ramesh Chandra}
\affiliation{Department of Physics, DSB Campus, Kumaun University, Nainital, 
India, 263 001 \\ email: {\tt reetikajoshi.ntl@gmail.com}} 
\pubyear{2018}
\jname{Astronomy in Focus, Volume 1} 
\editors{Piero Benvenuti, ed.}
\begin{document}

\maketitle
\vspace*{-0.3cm}
\begin{abstract}
We present here the observations of solar jets observed on April 04, 2017 from NOAA active region (AR) 
12644 using high temporal and spatial resolution AIA instrument. 
We have observed around twelve recurring jets during the whole day.
Magnetic flux emergence and cancellation have been observed 
at the jet location. The multi-band observations evidenced 
that these jets were triggered due to the magnetic reconnection at low coronal null--point.
\vspace*{-0.2cm}

\end{abstract}


\vspace{-0.7cm}
\section{Introduction}
Solar jets are small scale eruptions in the solar atmosphere observed at different wavelengths from
ground as well as from space (\cite[Shibata et al., 1992]{Shibata92},
\cite[Schmieder et al., 2013]{Schmieder13}, \cite[Joshi et al., 2017] {Joshi17}). 
These can be originated from quite as well as in active regions.
Several attempts have been made to understand the dynamics, kinematics and their magnetic topology.
It is believed now that the magnetic reconnection between the closed and open
magnetic field lines can be responsible for the origin of these
 phenomenon \cite[(Raoufi et al., 2017)]{Raoufi17}.
Magnetic flux emergence and cancellation have been observed at the jet locations
(for example see. \cite[Chandra et al., 2017]{Chandra17}). Magnetic topology of jet locations suggested that the null 
points,  Quasi Sepratix Layers (QSL), bald patches etc can be found at jet origin site (\cite[Mandrini et al., 1996]{Mandrini96}).
Numerical models have been proposed to shed light in this phenomenon 
(\cite[Pariat et al., 2009]{Pariat09}, 
\cite[Moreno-Insertis et al., 2008]{Moreno08}).
However it is debatable that what are the physical conditions
on the solar surface, responsible for their initiation.
 In this study, we present the multiwavelength study of recurrent jets observed in 
NOAA active region (AR) 12644 on April 04, 2017.
Section \ref{obs} presents observations and dynamics of the jets.
We summarize our results in section \ref{summary}.

\begin{figure}
\vspace*{1.0 cm}
\begin{center}
 \includegraphics[width=3in]{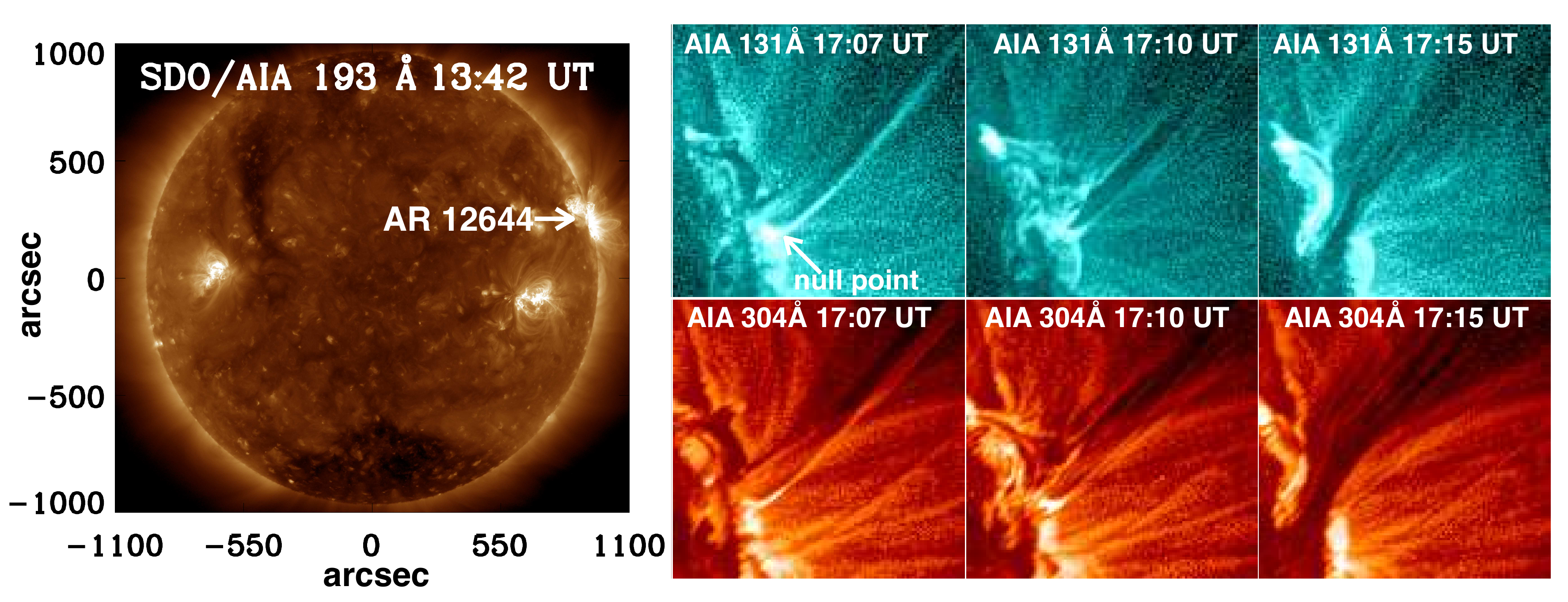} 
 \caption{Left panel: Location of NOAA AR 12644.
Right panel: An example of evolution of one jet in AIA 131\AA\ (first row) and 304\AA\ (second row) respectively. Location of the null--point
is marked by the white arrow.}   
\label{fig1}
\end{center}
\end{figure}

\begin{figure}
\begin{center}
 \includegraphics[width=3in]{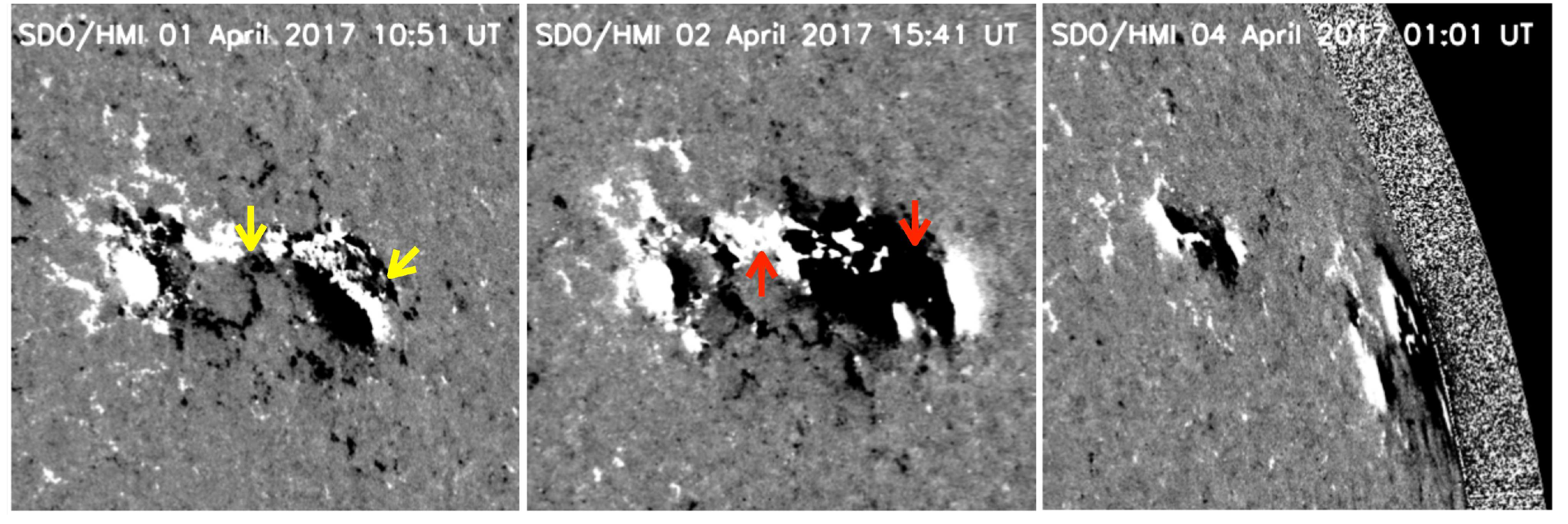}
 \caption{Evolution of magnetic field of NOAA AR 12644 from April 01, 2017 onwards.
Yellow and red arrows show the magnetic flux cancellation and emergence respectively.}
   \label{fig2}
\end{center}
\end{figure}

\vspace{-0.6cm}
\section{Observations and Results}
\label{obs}
For the present investigation, we have taken the data observed by
{\it Atmospheric Imaging Assembly} (AIA, \cite[Lemen et al., 2012]{Lemen12}) 
onboard {\it Solar Dynamics Observatory} (SDO, \cite[Pesnell, Thompson, \&
Chamberlin, 2012]{Pesnell12}). The magnetic field of the AR is 
explored by Heliospheric 
Magnetic Imager (HMI, \cite[Schou et al., 2012]{Schou12}) instrument. 

NOAA AR 12644 (N13W91) was one of the highly jet productive 
region on April 04, 2017.
It produced around twelve jets during the day in the north--west direction.
Figure \ref{fig1} (left) shows the location of AR in AIA 193 \AA\ .
The right panel of the Figure \ref{fig1} is an example of the evolution of 
one of the jets at 131 \AA\ 
and 304 \AA\ respectively. 
During the jets we have observed both cool and the hot structure. The jet structures
have reverse $``$Y" (or Eiffel tower) type structure.
Such Eiffel tower type morphology were also reported in the earlier studies
 for example: \cite[Torok et al., 2009]{Torok09}.
The Eiffel tower type structures evidenced the presence of magnetic null-point.
The possible location of the null-point is shown by the arrow in the first image 
of right panel in figure \ref{fig1}. We have measured the height of
observed null-point and it is $\sim$ 7 Mm. 
 We found the
average speed, height, and  lifetime of the jets as 100 kms$^{-1}$, 
60 Mm, and 10 min respectively.
For the magnetic causes of current jets, we have look at the 
magnetic field of the AR. The 
AR was at the west limb on April 04, 2017, therefore it is difficult to see its 
magnetic configuration due to large projection effect.
However, we track the evolution of the AR magnetic field from
April 01, 2017 onwards. Figure \ref{fig2} shows the evolution 
of magnetic field of AR.
We found the emergence of positive and negative flux at the jet location. Later
on this, newly emerged positive flux cancelled the existing negative flux, which confirms 
the reported findings that both the flux emergence and cancellation are 
responsible for the triggering of jets. 

\vspace{-0.6cm}
\section{Summary}
\label{summary}
We have reported here the preliminary results of twelve solar jets from
 NOAA AR 12644 on April 04, 2017. 
All the jets were clearly visible in all EUV channels which
 shows their multi--thermal nature. The morphology 
of  them indicate the presence of null--point type structure. Magnetic field evolution of the AR shows the magnetic flux emergence and cancellation. 
We believe that the presence of null--point and the magnetic 
emergence as well as cancellation at the jet's locations could be responsible for the production of these jets. 
In future, we plan to analyse them in more detail. \\

\vspace{-0.3cm}
{\bf Acknowledgments}
We acknowledge the financial support from IAU. RJ thanks Department of Science and Technology (DST),
New Delhi, India for an INSPIRE fellowship.

\end{document}